\documentclass[10pt,nofootinbib,superscriptaddress]{revtex4}
\usepackage[latin9]{inputenc}
\usepackage{helvet,amsmath,amssymb,graphicx,wasysym,epsfig,amsfonts,color,adjustbox,lipsum,datetime,slashed}
\usepackage[unicode=true, bookmarks=false, breaklinks=false,pdfborder={0 0 1},backref=section,colorlinks=false]{hyperref}
\usepackage[justification=raggedright, margin=5mm,font=sf,small,labelfont=bf, format=plain]{caption}
\definecolor{bluc}{cmyk}{1,1,0,0.1}

\begin{document}
\title{Purely Radiative Higgs Mass in Scale invariant models}
\author{Amine Ahriche $^{\href{https://orcid.org/0000-0003-0230-1774}{\includegraphics[width=2.5mm]{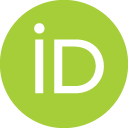}}}$}
\email{ahriche@sharjah.ac.ae}

\affiliation{Department of Applied Physics and Astronomy, University of Sharjah,
P.O. Box 27272 Sharjah, UAE.}
\affiliation{The Abdus Salam International Centre for Theoretical Physics, Strada
Costiera 11, I-34014, Trieste, Italy.}
\affiliation{Laboratoire de Physique des Particules et Physique Statistique, Ecole
Normale Superieure, BP 92 Vieux Kouba, DZ-16050 Algiers, Algeria.}
\begin{abstract}
In this work, we investigate the possibility of having scale invariant
(SI) standard model (SM) extensions, where the light CP-even scalar
matches the SM-like Higgs instead of being a light dilaton. After
deriving the required conditions for this scenario, we show that the
radiative corrections that give rise to the Higgs mass can trigger
the scalar mixing to the experimentally allowed values. In addition,
we discuss the constraints on the parameters space that makes the
CP-even scalars properties in a good agreement with all the recent
ATLAS and CMS measurements. We illustrate this scenario by considering
the SI-scotogenic model as an example, while imposing all the theoretical
and experimental constraints. We show that the model is viable and
leads to possible modifications of the di-Higgs signatures at current/future
with respect to the SM. 
\end{abstract}
\maketitle

\section{Introduction}

Despite the Higgs discovery~\cite{ATLAS:2012yve}, many questions
are still open within the standard model (SM), among them understanding
the origin of the Higgs mass. It is well known that in the SM, the
quadratic divergences appear in the radiative corrections to the Higgs
mass, which cause what is called the hierarchy problem. It is widely
believed that some possible solutions to the hierarchy problem come
with a classically conformally invariant action at a higher energy
scale that is below the gravity (Planck) scale, despite the fact that
any gravity theory is not conformally scale invariant due to the presence
of Planck mass as a dimensionful parameter~\cite{Meissner:2006zh}.
Since we do not know yet the gravity theory, one can assume the classical
SI invariance up to a scale below Planck by making the quadratic Higgs
mass term in the Lagrangian is set to zero ($\mu^{2}=0$). In this
setup, the electroweak symmetry breaking (EWSB) occurs via the so-called
dimensional transmutation, where the scale invariance symmetry is
broken at the quantum level~\cite{Coleman:1973jx}.

The SI symmetry breaking is associated by a pseudo-Goldstone boson
(PGB) that is strictly massless at tree-level; and acquires its mass
via the radiative corrections. This light scalar is called the ``dilaton''
scalar in the literature. The realization of the EWSB a la Coleman-Weinberg
within some SM popular extensions has been extensively discussed in
the literature (for example, see~\cite{Alexander-Nunneley:2010tyr}).
Although, many models are SI extended to address to the hierarchy
problem, in addition to other problems such as the dark matter (DM)
and the neutrino oscillation data~\cite{Foot:2007ay,Ahriche:2016cio}.
Here, we aim to investigate the case where the light PGB is the observed
$125~\mathrm{GeV}$ SM-like Higgs rather than a light dilaton. It
has been shown in the literature~\cite{Bellazzini:2012vz,Foot:2007as},
that the case of a dilaton-like Higgs, or a purely radiative mass
Higgs (PRMH) is possible, however, the requirements for such case
were not discussed, as well the agreement with the recent LHC measurements
relevant to the scalar sector.

In this work, we consider a generic SI model, where the SM is extended
by a real scalar singlet to assist the EWSB, in addition to new scalar
and fermionic field representations. In this general setup, we investigate
the EWSB and define the required conditions to have a PRMH case. Then,
we show how should this setup be in agreement with all the Higgs measurements~\cite{ATLAS:2016neq};
and the negative searches for a heavy resonance~\cite{ATLAS:2020zms,ATLAS:2020tlo,CMS:2021klu,ATLAS:2021nps,ATLAS:2021fet,ATLAS:2021ulo,ATLAS:2021jki}.
As will be shown later, the radiative corrections due to the interactions
of the new scalar and fermionic fields to the SM Higgs doublet and
the real singlet; could play a key role. As they give rise to the
Higgs mass, they are also responsible to adjust the scalar mixing
to be in agreement with the recent constraints; and fully control
some triple scalar couplings that can be directly probed through the
di-Higgs production at both LHC and ILC. In order to illustrate our
discussion, we consider a phenomenologically rich SI model~\cite{Ahriche:2016cio}
as an example.

This work is organized as follow: in section~\ref{sec:EWSB}, we discuss the EWSB
and deduce the required conditions for a PRMH case is section~\ref{sec:PRHM}.
Then, the experimental constraints and the predictions at colliders
are investigated in section~\ref{sec:Const}. Section~\ref{sec:ex} is devoted for an illustrative
example and our conclusions are summarized in section~\ref{sec:conc}.

\section{The EWBS in SI Models\label{sec:EWSB}}

The classical scale invariance symmetry enforces the action to be
invariant under the conformal transformation\footnote{Here, $\kappa$ and $a$ are real numbers, where $a=1$ for bosons
and $a=3/2$ for fermions.} $\Phi_{i}(x^{\mu})\rightarrow e^{a~\kappa}\,\Phi_{i}(e^{\kappa}\,x^{\mu})$,
which implies the vanishing of the scalar quadratic and the fermionic
mass terms in the Lagrangian density. Then, for a model with many
scalar representations, the scalar potential can be written in the
general form 
\begin{equation}
V=\sum_{i,j,k,l}\lambda_{ijkl}\Phi_{i}\Phi_{j}\Phi_{k}\Phi_{l},
\end{equation}
where the couplings $\lambda_{ijkl}$ are not vanishing due to the
symmetries that are assigned to the model. Generally, most of the
SI models in the literature include the SM Higgs doublet $\mathcal{H}^{T}=\Big(\chi^{+},~[h+i~\chi^{0}]/\sqrt{2}\Big)$,
a real scalar singlet $\phi$ to assists the EWSB in addition to other
bosonic and fermionic representations with different multiplicities.
After the EWSB, the CP-even neutral scalars acquire their VEV's as
\begin{equation}
\mathcal{H}\rightarrow\frac{\upsilon+h}{\sqrt{2}}\begin{pmatrix}0\\
1
\end{pmatrix},\,\,\,\phi\rightarrow x+\phi,\label{VEV}
\end{equation}
and give masses to all the model fields. Then, we get two CP-even
eigenstates $h_{1,2}$ ($m_{1}<m_{2}$) via a rotation with the angle
$\alpha$ in the basis \{$h,\phi$\}, where one of the eigenstates
must match the SM-like Higgs with the measured mass $m_{h}=125.18\,\mathrm{GeV}$.
In the literature, the heavier eigenstate $h_{2}=H$ corresponds to
the SM-like Higgs and $h_{1}=D$ is the dilaton scalar, that is strictly
massless at tree-level and acquires its mass via the radiative corrections.
The other case corresponds to a purely radiative mass Higgs (PRMH)
scenario, i.e., $h_{1}=H$ and $h_{2}=S$ would be a heavy CP-even
scalar. The aim of this work is to investigate the viability of the
PRMH scenario and to show possible interesting signatures at colliders.

In order to achieve the EWSB, one has to consider the radiative corrections
to the scalar potential. The one-loop effective scalar potential can
be written in function of the CP-even scalar fields as 
\begin{eqnarray}
V^{1-\ell}(h,\phi) & = & \frac{1}{24}(\lambda_{h}+\delta\lambda_{h})h^{4}+\frac{1}{24}(\lambda_{\phi}+\delta\lambda_{\phi})\phi^{4}+\frac{1}{4}(\omega+\delta\omega)h^{2}\,\phi^{2}+\sum_{i}n_{i}G\Big(m_{i}^{2}\Big),\label{eq:V1l}
\end{eqnarray}
where $\delta\lambda_{h},\,\delta\lambda_{\phi}$ and $\delta\omega$
are the counter-terms, $n_{i}$ and $m_{i}^{2}\equiv m_{i}^{2}(h,\phi)$
are the field multiplicities and field dependent squared masses. Here,
the function $G(r_{i})=\frac{r_{i}^{2}}{64\pi^{2}}\Big(\log\frac{r_{i}}{\Lambda^{2}}-c_{i}\Big)$
is defined a la the $\overline{DR}$ scheme ($c_{i}=3/2$) and $\Lambda$
is the renormalization scale. The appearance of the dimensionful parameter
$\Lambda$ shows the scale invariance is broken, and when studying
the phenomenology of the model, such as the physics at the colliders,
it should take a value of the electroweak scale order like $\Lambda=m_{h}$.

Including the CTs is mandatory to in order cancel the divergences
that appear from the one-loop corrections, and therefore regularize
the theory. The way these infinities are absorbed by the CTs depends
on the definition of the renormalized parameters, i.e., on the choice
of the renormalization conditions. Here, in our work I adopted a modified
version of the $\bar{DR}$ scheme, where the choice of the CTs (more
precisely their cut-off independent parts) makes the values of the
masses and mixing at tree-level and one-loop having identical values
at the vacuum $\{h=\upsilon,\phi=x\}$. In other words, the CTs should
be derived from the three conditions $\left.\partial V^{1-\ell}/\partial h\right|_{h=\upsilon,\phi=x}=\left.\partial V^{1-\ell}/\partial\phi\right|_{h=\upsilon,\phi=x}=0$
and the Higgs mass $m_{1,2}^{2\,(1-\ell)}=m_{h}^{2}$.

Using the tadpole conditions, the one-loop scalar squared mass matrix
in the basis \{$h,\phi$\}, can be written in function of $\delta\omega$,
as 
\begin{equation}
\textrm{M}^{2}=\left[\frac{m_{h}^{2}}{\upsilon^{2}+x^{2}}-\delta\omega\right]\left(\begin{array}{cc}
x^{2} & -\upsilon x\\
-\upsilon x & \upsilon^{2}
\end{array}\right)+m_{h}^{2}\left(\begin{array}{cc}
a & c\\
c & b
\end{array}\right),\label{M2}
\end{equation}
and the one-loop contributions to the Higgs/dilaton mass are characterized
by the dimensionless parameters 
\begin{align}
a & =\frac{1}{m_{h}^{2}}\sum_{i}n_{i}\left[\left(\partial_{h,h}m_{i}^{2}-\frac{3}{\upsilon}\partial_{h}m_{i}^{2}\right)G^{\prime}(m_{i}^{2})+\left(\partial_{h}m_{i}^{2}\right)^{2}G^{\prime\prime}(m_{i}^{2})\right]_{h=\upsilon,\phi=x},\\
b & =\frac{1}{m_{h}^{2}}\sum_{i}n_{i}\left[\left(\partial_{\phi,\phi}m_{i}^{2}-\frac{3}{x}\partial_{\phi}m_{i}^{2}\right)G^{\prime}(m_{i}^{2})+\left(\partial_{\phi}m_{i}^{2}\right)^{2}G^{\prime\prime}(m_{i}^{2})\right]_{h=\upsilon,\phi=x},\\
c & =\frac{1}{m_{h}^{2}}\sum_{i}n_{i}\left[\Big(\partial_{h,\phi}m_{i}^{2}\Big)G^{\prime}(m_{i}^{2})+(\partial_{h}m_{i}^{2})(\partial_{\phi}m_{i}^{2})G^{\prime\prime}(m_{i}^{2})\right]_{h=\upsilon,\phi=x},\label{ABC}
\end{align}
where $G^{\prime}(r)=\partial G(r)/\partial r$, $G^{\prime\prime}(r)=\partial^{2}G(r)/\partial r^{2}$,
$\partial_{x}=\frac{\partial}{\partial x}$ and $\partial_{x,y}=\frac{\partial^{2}}{\partial x\partial y}$.
In order to find the value of the counter-term $\delta\omega$, we
require the measured Higgs mass to match one of the eigenmasses, i.e.,
$2m_{h}^{2}=\textrm{M}_{11}^{2}+\textrm{M}_{22}^{2}\pm\{(\textrm{M}_{22}^{2}-\textrm{M}_{11}^{2})^{2}+4(\textrm{M}_{12}^{2})^{2}\}^{1/2}$.
Both cases give the same value for $\delta\omega$, 
\begin{align}
\delta\omega & =\frac{m_{h}^{2}}{\upsilon^{2}+x^{2}}\frac{(ab-c^{2})(\upsilon^{2}+x^{2})-ax^{2}-b\upsilon^{2}+2\,c\upsilon x}{a\upsilon^{2}+bx^{2}+2\,c\upsilon x-\upsilon^{2}-x^{2}}.\label{dw}
\end{align}

Numerically, the counter-terms $\delta\omega$, $\delta\lambda_{h}$
and/or $\delta\lambda_{\phi}$ may acquire large values, especially
for large singlet VEV $x$, non-negligible dimensionless couplings
and/or large fields multiplicities. To avoid such naturalness, one
has to impose the perturbativity constraints at one-loop level. This
can be achieved by considering the one-loop quartic couplings 
\begin{equation}
\lambda_{h,\phi}^{1-\ell},\,|\omega^{1-\ell}|<4\,\pi,\label{pert}
\end{equation}
where these one-loop couplings are defined as the $4^{th}$ derivatives
of the effective potential (\ref{eq:V1l}) at the vacuum \{$h=\upsilon,~\phi=x$\}.
Although, there is no need to impose the vacuum stability conditions
at tree-level $\min[\lambda_{h},~\lambda_{\phi},~3\omega+\{\lambda_{h}\lambda_{\phi}\}^{1/2}]>0$
or at one-loop $\min[\lambda_{h}^{1-\ell},~\lambda_{\phi}^{1-\ell},~3\omega^{1-\ell}+\{\lambda_{h}^{1-\ell}\lambda_{\phi}^{1-\ell}\}^{1/2}]>0$,
since the leading term in the effective potential (\ref{eq:V1l})
is $\varphi^{4}\log\varphi$ rather than $\varphi^{4}$, where $\varphi$
stands for any direction in the plan \{$h,\phi$\}. Therefore, the
one-loop conditions of the vacuum stability come from the coefficients
positivity of the terms $\varphi^{4}\log\varphi$ in the effective
potential. In other words, we must have 
\begin{equation}
\sum_{i}n_{i}m_{i}^{2}(h=\upsilon,\phi=0)>0,~~~~\sum_{i}n_{i}m_{i}^{2}(h=0,\phi=x)>0,\label{VS}
\end{equation}
as the one-loop vacuum stability conditions~\cite{Soualah:2021xbn}.
Concerning the SI breaking scale $\Lambda_{0}$, one has to estimate
the RGE solution for quartic couplings; and estimate the running up
to higher scale, much higher than $\Lambda=m_{h}$, then, $\Lambda_{0}$
can be defined as the scale where of the perturbativity and/or the
vacuum stability conditions get broken. This depends on the model
field content, multiplicities, and couplings. Such analysis about
the vacuum structure at higher energy scales within the SI-scotogenic
model is under investigation~\cite{ahriche}.

\section{The Purely Radiative Mass Higgs\label{sec:PRHM}}

After the EWSB, we obtain two CP-even eigenstates in the PRMH scenario
as 
\begin{equation}
\begin{pmatrix}H\\
S
\end{pmatrix}=\begin{pmatrix}c_{\alpha}~-s_{\alpha}\\
s_{\alpha}~c_{\alpha}
\end{pmatrix}\begin{pmatrix}h\\
\phi
\end{pmatrix},\label{scMix}
\end{equation}
where $c_{\alpha}=\cos\alpha,\,s_{\alpha}=\sin\alpha$ , $H$ denotes
the 125 $\mathrm{GeV}$ Higgs, $S$ is the new heavy scalar and $\alpha$
is the scalar mixing angle, that is defined by 
\begin{equation}
\tan2\alpha=2\textrm{M}_{12}^{2}/\Big[\textrm{M}_{22}^{2}-\textrm{M}_{11}^{2}\Big],\label{eq:mix}
\end{equation}
with $\textrm{M}_{ij}^{2}$ are the elements of (\ref{M2}).

Depending on the model free parameters (the singlet VEV $x$ and the
fields couplings to the real scalar singlet and the Higgs doublet),
the observed $125~\mathrm{GeV}$ SM-like Higgs could match the heavier
(light dilaton case) or the lighter (PRHM case) CP-even eigenstate.
The light dilaton case is possible only if $m_{h}^{2}<\textrm{M}_{11}^{2}+\textrm{M}_{22}^{2}<2\,m_{h}^{2}$
that can be translated into 
\begin{equation}
\delta\omega(x^{2}+\upsilon^{2})/m_{h}^{2}<a+b<1+\delta\omega(x^{2}+\upsilon^{2})/m_{h}^{2},\label{eq:Dil}
\end{equation}
This condition (\ref{eq:Dil}) ensures that the dilaton squared mass
is positive and smaller than the Higgs one. The PRMH scenario is possible
if $\textrm{M}_{11}^{2}+\textrm{M}_{22}^{2}>2\,m_{h}^{2}$, which
leads to 
\begin{equation}
a+b>1+\delta\omega(x^{2}+\upsilon^{2})/m_{h}^{2}.\label{eq:Cond}
\end{equation}

\begin{figure}[h]
\includegraphics[width=0.33\textwidth]{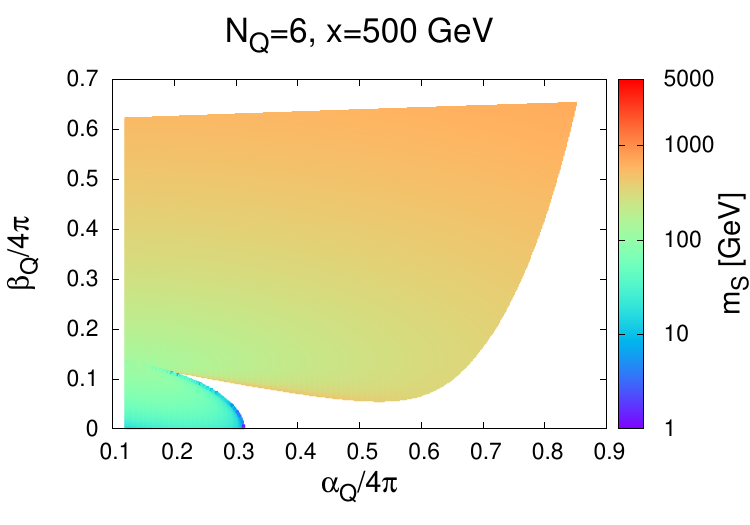}~\includegraphics[width=0.33\textwidth]{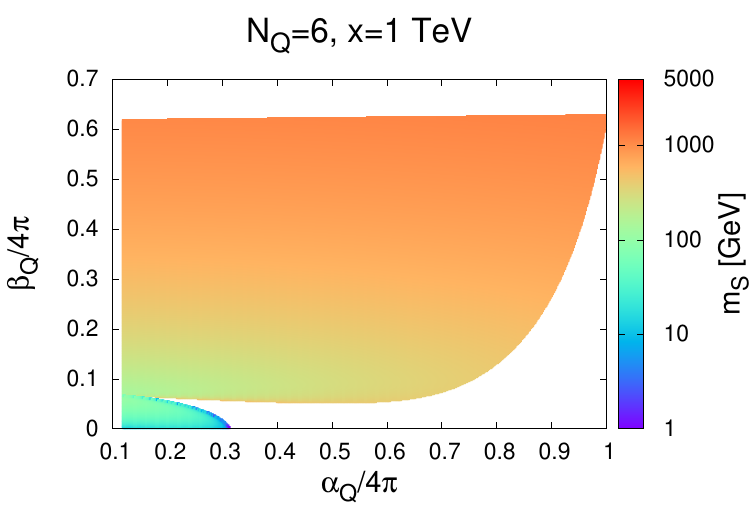}~\includegraphics[width=0.33\textwidth]{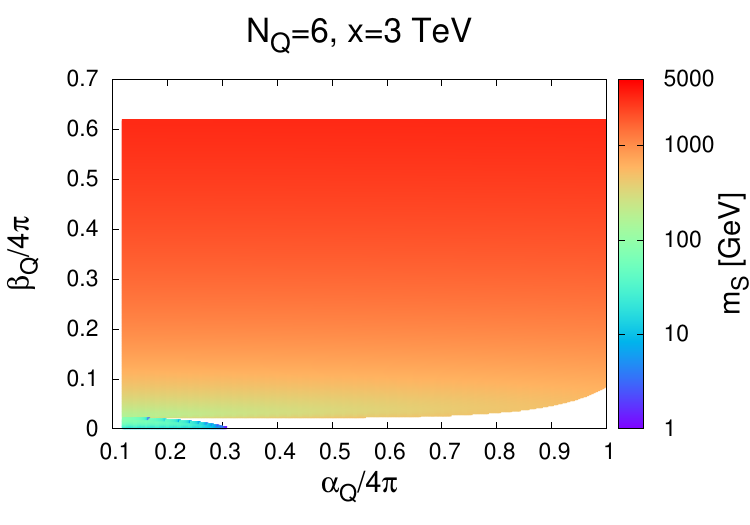}\\
 \includegraphics[width=0.33\textwidth]{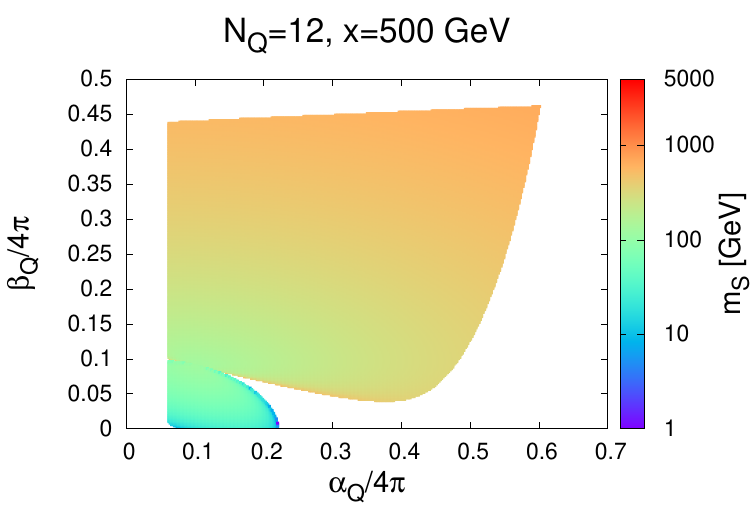}~\includegraphics[width=0.33\textwidth]{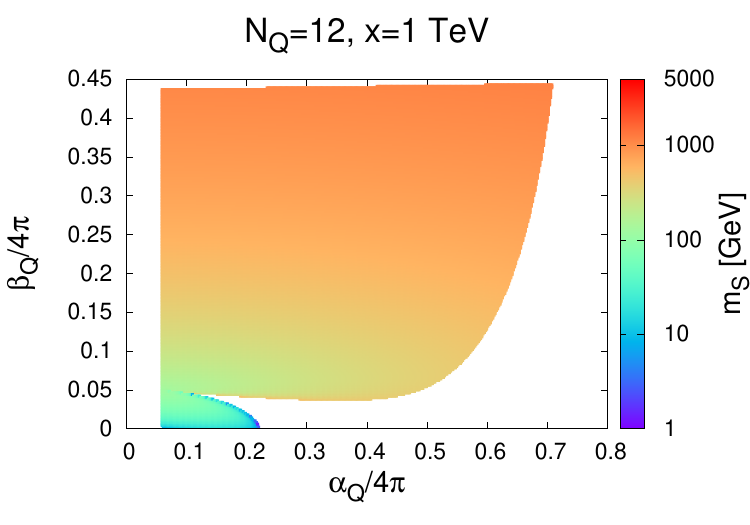}~\includegraphics[width=0.33\textwidth]{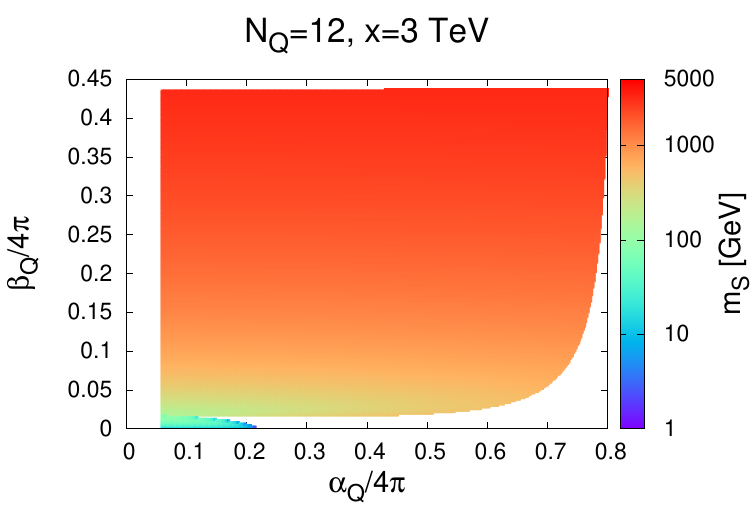}\\
 \includegraphics[width=0.33\textwidth]{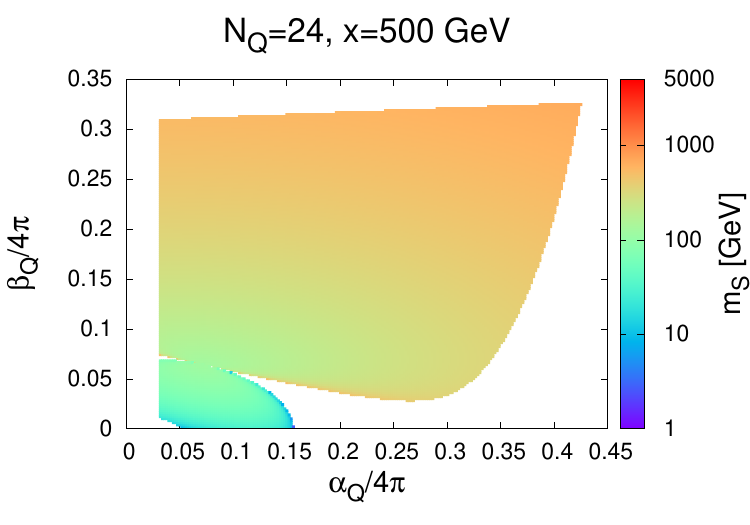}~\includegraphics[width=0.33\textwidth]{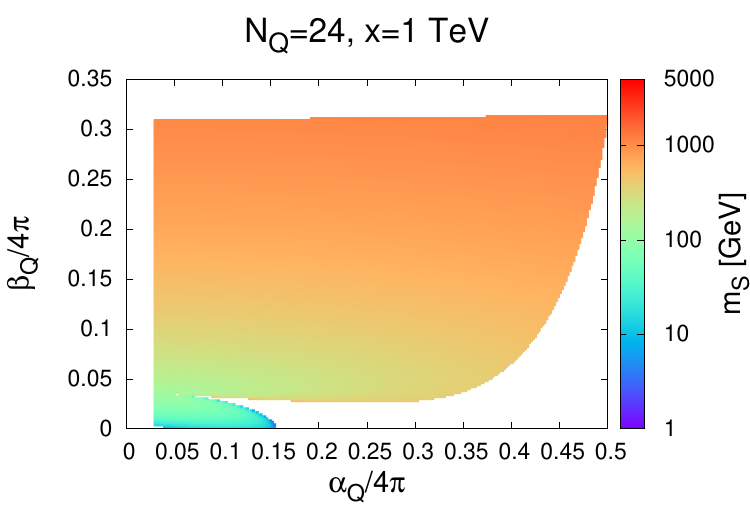}~\includegraphics[width=0.33\textwidth]{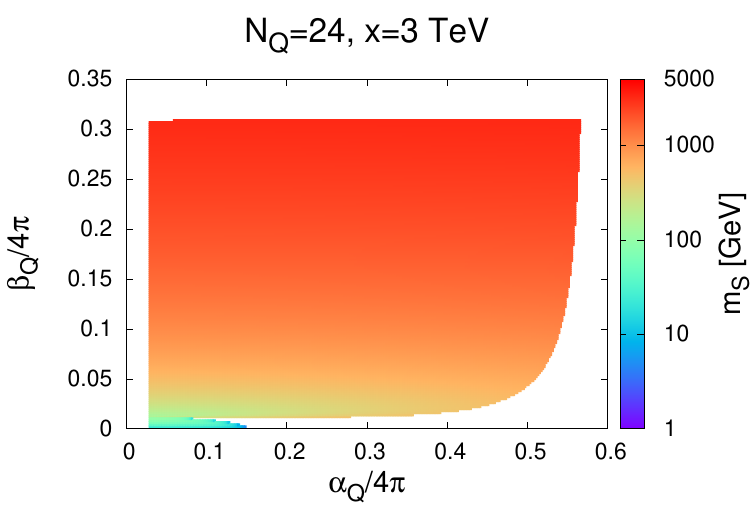}
\caption{The couplings $\alpha_{Q}$ and $\beta_{Q}$, where the palette shows
the extra scalar mass $m_{Q}$, and the one-loop perturbativity (\ref{pert});
and vacuum stability (\ref{VS}) constraints are considered. These
upper larger (lower smaller) island corresponds to the PRMH (light
dilaton) scenario.}
\label{mS} 
\end{figure}

\begin{figure}[h]
\includegraphics[width=0.33\textwidth]{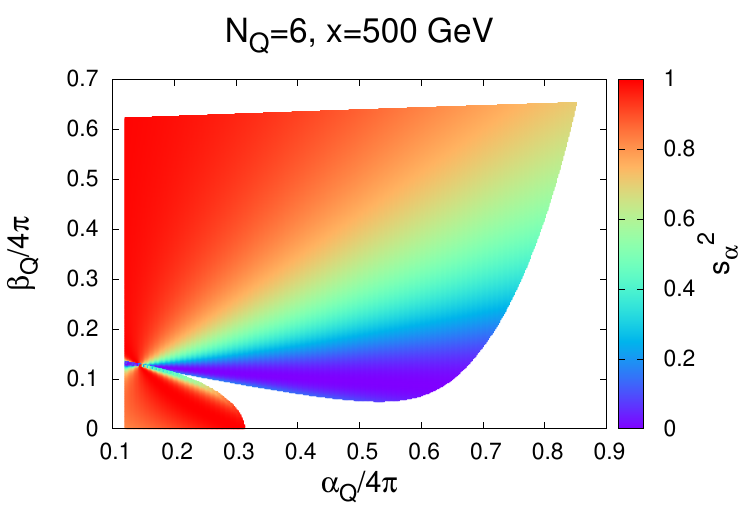}~\includegraphics[width=0.33\textwidth]{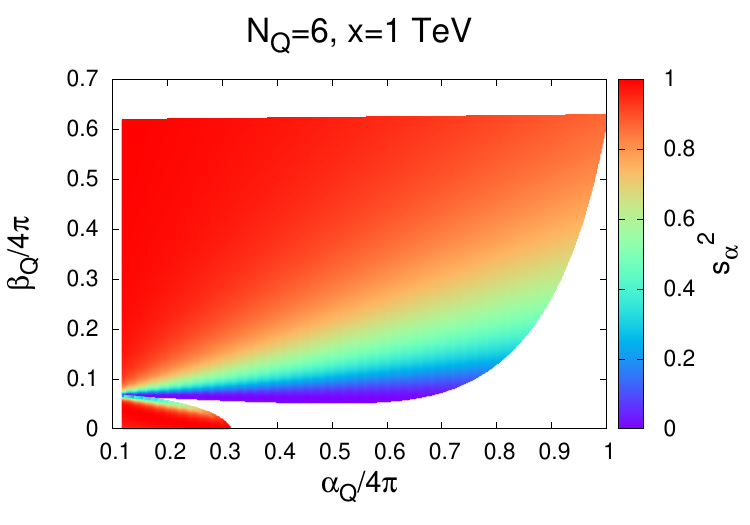}~\includegraphics[width=0.33\textwidth]{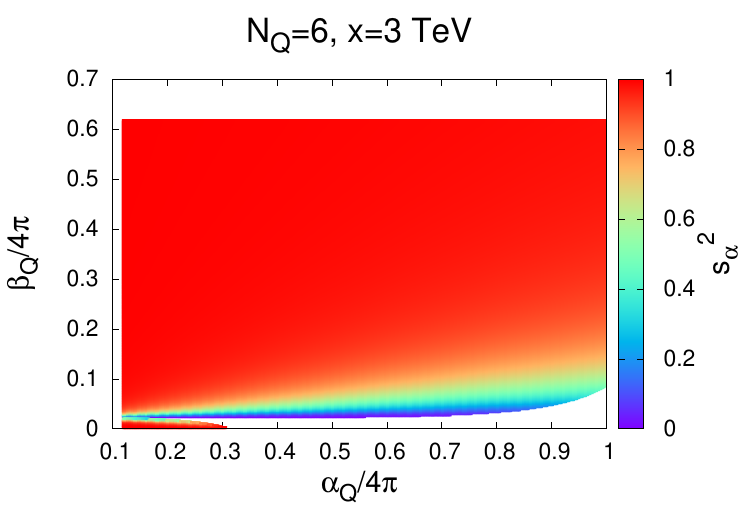}\\
 \includegraphics[width=0.33\textwidth]{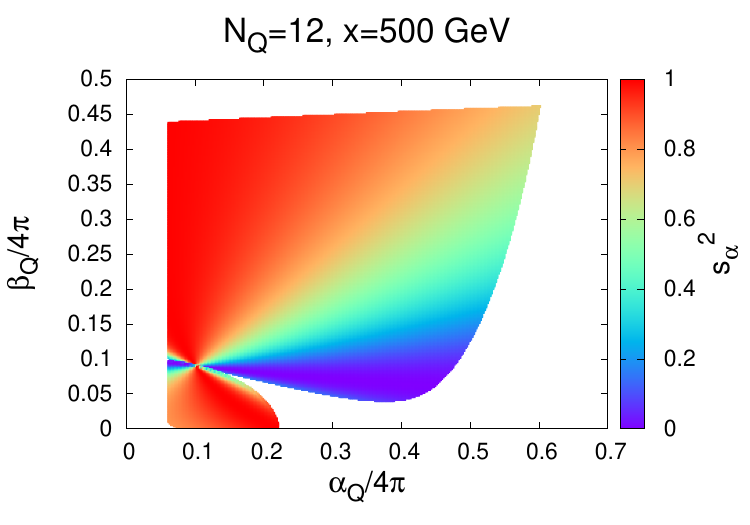}~\includegraphics[width=0.33\textwidth]{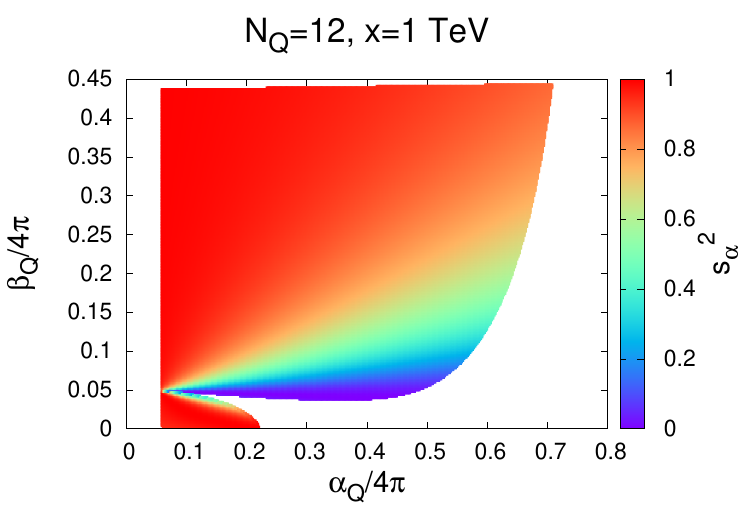}~\includegraphics[width=0.33\textwidth]{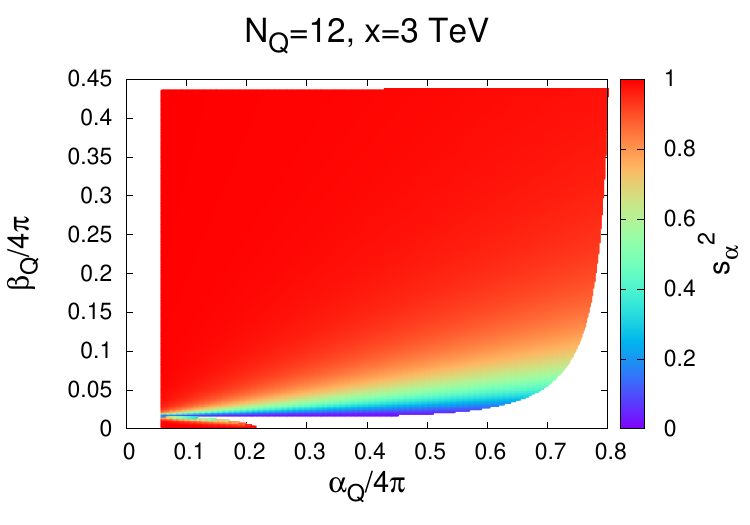}\\
 \includegraphics[width=0.33\textwidth]{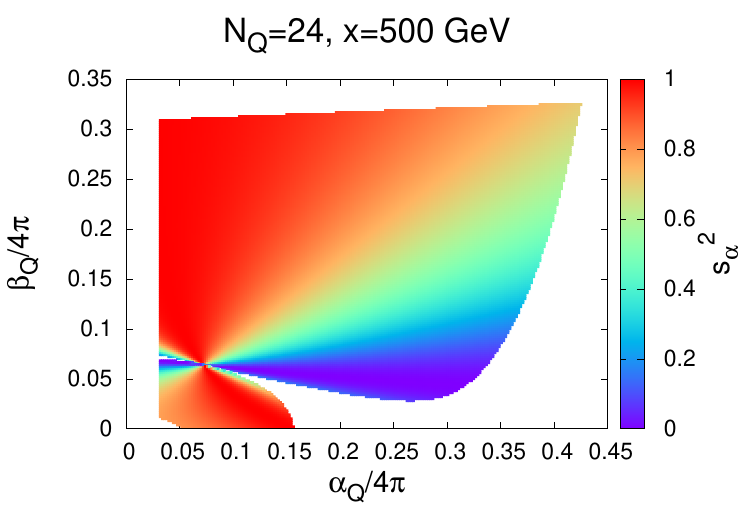}~\includegraphics[width=0.33\textwidth]{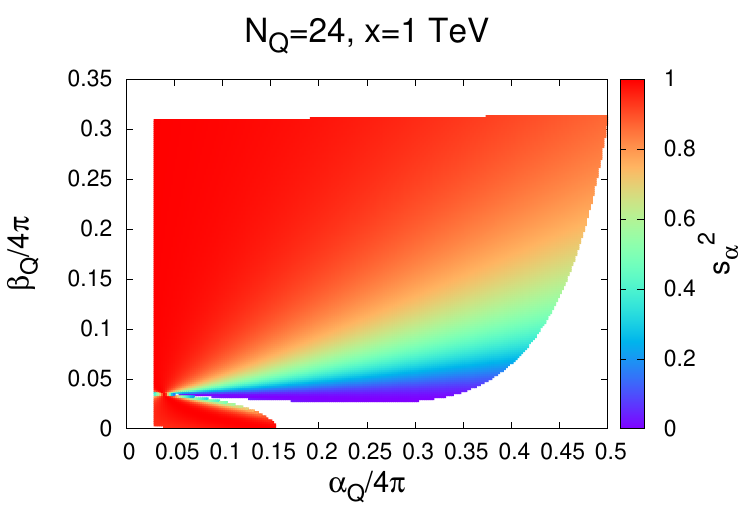}~\includegraphics[width=0.33\textwidth]{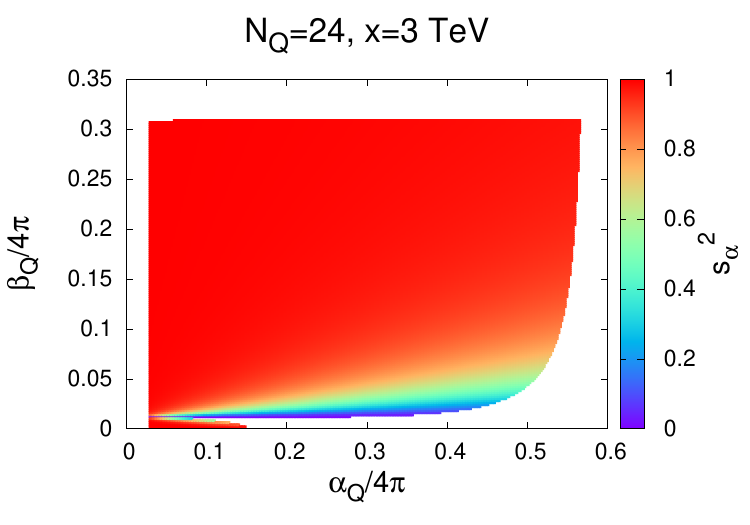}
\caption{The couplings $\alpha_{Q}$ and $\beta_{Q}$, where the palette shows
the scalar mixing $s_{\alpha}^{2}$, and the one-loop perturbativity
(\ref{pert}); and vacuum stability (\ref{VS}) constraints are considered.
These upper larger (lower smaller) island corresponds to the PRMH
(light dilaton) scenario.}
\label{sa} 
\end{figure}

In order to have an idea about the quantum corrections that lead to
the PRMH scenario (i.e., fulfilling the condition (\ref{eq:Cond}))
compared to the light dilaton one, we consider a toy model where the
SM is extended by a scalar singlet (\ref{VEV}) to assist the EWSB;
and another singlet scalar $Q$ with multiplicity $N_{Q}$ and the
squared mass $m_{Q}^{2}=\frac{1}{2}(\alpha_{Q}\upsilon^{2}+\beta_{Q}x^{2})$.
Clearly, the quantum corrections effect should be proportional to
the field multiplicity $N_{Q}$, the couplings ($\alpha_{Q},\,\beta_{Q}$)
to $\mathcal{H}$ and $\phi$ and/or the singlet VEV $x$. To confirm
this, we show in Fig.~\ref{mS} (Fig.~\ref{sa}), the parameter
space ($\alpha_{Q},~\beta_{Q}$) for both light dilaton and PRHM cases,
where the palette shows $m_{Q}$ (the mixing $s_{\alpha}$) for different
values of the multiplicity $N_{Q}=6,12,24$; and the singlet VEV $x=500\,\mathrm{GeV},~1\,\mathrm{TeV},~3\,\mathrm{TeV}$.
These figures are produced by taking into account the perturbativity
one-loop constraints (\ref{pert}), in addition to the vacuum stability
(\ref{VS}).

From Fig.~\ref{mS}, one sees that in all panels there exist two
islands; a lower smaller island and a larger upper one, which corresponds
to the light dilaton and PRHM cases, respectively. Clearly, the parameter
space in the PRHM case is much larger than the light dilaton case.
Indeed, this is easy to understand since the radiative corrections
(i.e., values of $N,~,\alpha_{Q},~and~\beta_{Q}$) that are required
to achieve the EWSB and make the light CP-even eigenstate matching
the observed SM-like Higgs; should be much larger than case of breaking
the EW symmetry and give a tiny mass to the dilaton.

In both Fig.~\ref{mS} and Fig.~\ref{sa}, the shape of the parameters
space for different values of the singlet VEV ($x$) and the new scalar
multiplicity ($N_{Q}$) is dictated by many constraints such as the
positivity of $m_{Q}^{2}>0$, the one-loop perturbativity (\ref{pert});
the vacuum stability conditions (\ref{VS}); and the definition of
both light dilaton and PRMH scenarios. One has to mention that the
two region are connected in a point at least, which corresponds to
the case of two degenerate scalars at the mass $m=125~\mathrm{GeV}$.
This twin Higgs scenario could be of great interest~\cite{Heikinheimo:2013cua}.

From Fig.~\ref{mS}, one learns that the condition (\ref{eq:Cond})
can be fulfilled for small couplings ($\alpha_{Q}$, $\beta_{Q}$)
and small masses. However, for larger $m_{Q}$ values, i.e., by making
the singlet VEV ($x$) larger, the PRMH scenario can be achieved for
larger values of the couplings ($\alpha_{Q}$, $\beta_{Q}$). It is
clear that pushing the singlet VEV to higher values leads to the decoupling
limit. From Fig.~\ref{sa}, one notices that the light green color
corresponds to small scalar mixing values, that is in agreement with
the experimental constraints as we will see in the next section. Obviously,
it clear that for larger singlet VEV and multiplicity values, the
viable parameters space is larger for the PRHM scenario. Here, one
has to mention that in realistic models where many fermionic and scalar
degrees of freedom and added to the SM, there will be more parameters,
more freedom and more theoretical and experimental constraints, as
will be seen in section V. In what follows, we will be interested
in PRHM scenario.

\section{Constraints \& Predictions\label{sec:Const}}

Both ATLAS and CMS measurements at $\sqrt{s}=7+8\,\mathrm{TeV}$ reported
the total Higgs signal strength modifier to be $\mu_{{\rm tot}}=c_{\alpha}^{2}\times(1-\mathcal{B}_{BSM})\geq0.89$
at 95~\% CL~\cite{ATLAS:2016neq}, which implies $s_{\alpha}^{2}\leq0.11$
in the absence of invisible and undetermined Higgs decay ($\mathcal{B}_{BSM}=0$).
Since the tree-level scalar mixing in the PRMH scenario is defined
by $s_{\alpha}^{(0)}=-x/(\upsilon^{2}+x^{2})^{1/2}$, the bound ($(s_{\alpha}^{(0)})^{2}\leq0.11$)
leads to contradictory values for the singlet VEV $x\leq86.6~\mathrm{GeV}$.
Then, if one writes $s_{\alpha}=(1+\Delta_{\sin\alpha})s_{\alpha}^{(0)}$,
the radiative corrections to the mixing must be large and negative
\begin{equation}
|1+\Delta_{\sin\alpha}|<\sqrt{0.11}(1+\upsilon^{2}/x^{2})^{1/2},\label{cond}
\end{equation}
in order to have a viable PRMH scenario for natural values of the
singlet VEV $x>\upsilon$. Therefore, the radiative effects (quantified
by the fields multiplicities and couplings to the Higgs doublet and
to the real singlet) must push the light CP-even scalar mass to match
$m_{h}$ and give a large negative contribution to the scalar mixing
sine ($s_{\alpha}$) at the same time to have a viable SM-like Higgs
in the PRMH scenario. The radiative corrections to the scalar mixing
(\ref{cond}) in this scenario are so constrained with respect to
the light dilaton scenario, since the tree-level mixing $s_{\alpha}^{(0)}=\upsilon/(\upsilon^{2}+x^{2})^{1/2}$
is naturally small, and therefore allows large (positive or negative)
radiative corrections $|\Delta_{\sin\alpha}|\leq1500\%$~\cite{Soualah:2021xbn}.

In the PRMH scenario, in addition to the constraints on the Higgs
due to the Higgs signal strength modifiers, the di-photon, invisible
and undetermined decays and the total Higgs decay width, the new heavy
CP-even scalar $S$ is a subject of constraints from many negative
searches at the LHC. Since the CP-even field of the Higgs doublet
can be written as $h=H~c_{\alpha}+S~s_{\alpha}$, then the scalar
S has the same SM-like Higgs couplings to the SM fermions and gauge
bosons scaled either by $s_{\alpha}$ or $s_{\alpha}^{2}$. Hence,
it decays to all the SM Higgs final states, di-Higgs final state or
via other invisible or undetermined channels according to the model
field content. This allows different search types among them looking
for a heavy CP-even resonance in the channels of pair of leptons,
jets or gauge bosons $pp\rightarrow S\rightarrow\ell\ell,jj,VV$;
and the search of resonant di-Higgs production $pp\rightarrow S\rightarrow HH$.
For the first type, we consider the recent ATLAS analysis at $13~\mathrm{TeV}$
with 139 $\mathrm{fb}^{-1}$ $pp\rightarrow S\rightarrow\tau\tau$~\cite{ATLAS:2020zms},
and $pp\rightarrow S\rightarrow ZZ$ via the channels $\ell\ell\ell\ell$
and $\ell\ell\nu\nu$~\cite{ATLAS:2020tlo}, in addition to the CMS
analysis at $13~\mathrm{TeV}$ with 137 $\mathrm{fb}^{-1}$ $pp\rightarrow S\rightarrow WW$~\cite{CMS:2021klu}.
For the second type, we consider the recent ATLAS combination~\cite{ATLAS:2021nps},
that includes the analyses at $13~\mathrm{TeV}$ with 139 $\mathrm{fb}^{-1}$
via the channels $HH\rightarrow b\bar{b}\tau\tau$~\cite{ATLAS:2021fet},
$HH\rightarrow b\bar{b}b\bar{b}$~\cite{ATLAS:2021ulo} and $HH\rightarrow b\bar{b}\gamma\gamma$~\cite{ATLAS:2021jki}.

In all SI extensions of the SM where the EWSB is assisted by a singlet
scalar $\phi$, the triple scalar couplings $\lambda_{HHH}$ and $\lambda_{HHS}$
are strictly vanishing at tree-level. Therefore, any process that
is sensitive to these scalar triple couplings (like $pp\rightarrow HH@LHC14$
and $e^{-}e^{+}\rightarrow ZHH@ILC500$ as examples) would be fully
triggered by radiative effects. Since the radiative contributions
to the scalar mixing ($\Delta_{s_{\alpha}}$) are expected to be large
and negative, one has to consider the one-loop scalar mixing to get
a precise estimation for these triple couplings $\lambda_{HHH}$ and
$\lambda_{HHS}$. By considering the one-loop scalar mixing, significantly
improved (re-summed) values for these couplings could be obtained
as the third derivatives of the one-loop effective potential (\ref{eq:V1l})~\cite{Ahriche:2013vqa}.
The details are shown in appendix~\ref{app}, the triple couplings
$\lambda_{HHH}$ and $\lambda_{HHS}$ are estimated.

In this setup, one can classify the Feynman diagrams describing the
processes $pp\rightarrow HH@LHC14$ and $e^{-}e^{+}\rightarrow ZHH@ILC500$
in diagrams with and without the triple scalar couplings ($\lambda_{HHH,HHS}$).
Therefore, the cross section has the different contributions (1) that
involves only $\lambda_{HHH,HHS}$ ($\sigma_{\lambda}$) diagrams,
(2) with a pure gauge couplings contribution ($\sigma_{G}$); and
(3) the interference contribution ($\sigma_{G\lambda}$). This makes
the cross section in both processes modified with respect to the SM
value as 
\begin{equation}
R(f)=\frac{\sigma(f)-\sigma_{SM}(f)}{\sigma_{SM}(f)}=\frac{\xi_{1}\sigma_{G}+\xi_{2}\sigma_{\lambda}+\xi_{3}\sigma_{G\lambda}}{\sigma_{G}+\sigma_{\lambda}+\sigma_{G\lambda}}-1,\label{HH}
\end{equation}
with $f\equiv pp\rightarrow HH@LHC14$ and $e^{-}e^{+}\rightarrow ZHH@ILC500$.
For the process $f\equiv pp\rightarrow HH@LHC14$, we have $\sigma_{G}\equiv\sigma_{\Square}=70.1\,\mathrm{fb}$,~$\sigma_{\lambda}\equiv\sigma_{\triangle}=9.66\,\mathrm{fb}$
and $\sigma_{G\lambda}\equiv\sigma_{\triangle\Square}=-49.9\,\mathrm{fb}$
are the box, triangle and interference contributions to the total
cross section, respectively~\cite{Spira:1995mt}. Using MadGraph~\cite{Alwall:2011uj},
we find $\sigma_{G}=0.0837~\mathrm{fb},~\sigma_{\lambda}=0.01565~\mathrm{fb}$
and $\sigma_{G\lambda}=0.05685~\mathrm{fb}$ for the process $e^{-}e^{+}\rightarrow ZHH@ILC500$.
The coefficients $\xi_{i}$ are given at the CM energy $\sqrt{s}$
by~\cite{Baouche:2021wwa} 
\begin{align}
\xi_{1}=c_{\alpha}^{4},\,\xi_{2} & =\left|\mathcal{P}\right|^{2},\,\xi_{3}=c_{\alpha}^{2}\Re\left(\mathcal{P}\right),\,\mathcal{P}=c_{\alpha}\frac{\lambda_{HHH}}{\lambda_{hhh}^{SM}}+s_{\alpha}\frac{\lambda_{HHS}}{\lambda_{hhh}^{SM}}\frac{s-m_{h}^{2}+im_{h}\Gamma_{h}}{s-m_{S}^{2}+im_{S}\Gamma_{S}}\label{eq:coef}
\end{align}
with $\Gamma_{h}=4.2\,\mathrm{MeV}$ is the measured Higgs total decay
width, $\Gamma_{S}$ is the estimated heavy scalar total decay width
and $\lambda_{hhh}^{SM}$ is the SM Higgs triple coupling that is
estimated as in~\cite{Kanemura:2002vm}.

\section{Illustrative Example\label{sec:ex}}

In order to illustrate this discussion, we consider a phenomenologically
rich SI example, the SI-scotogenic model~\cite{Ahriche:2016cio},
where the SM is extended by one inert doublet scalar, $S^{T}\equiv\Big(S^{\pm}\,,\,[S^{0}+iA^{0}]/\sqrt{2}\Big)$,
three singlet Majorana fermions $N_{i}$, and one real neutral singlet
scalar $\phi$. The model is assigned by a global $Z_{2}$ symmetry
$\{S,\,N_{i}\}\rightarrow\{-S,\,-N_{i}\}$, where all other fields
being $Z_{2}$-even. This global symmetry makes the lightest $Z_{2}$-odd
field as a stable DM candidate (which we take $N_{1}$ in our example).
One easily constructs the effective potential (\ref{eq:V1l}) for
this model by deriving the field dependent masses through the relevant
parts of the SI invariant Lagrangian density 
\begin{align}
-\mathcal{L}\supset & =\frac{1}{2}y_{i}\phi\overline{N_{i}^{c}}\,N_{i}+\frac{1}{6}\lambda_{h}(\left|\mathcal{H}\right|^{2})^{2}+\frac{\lambda_{\phi}}{24}\phi^{2}+\frac{\lambda_{S}}{2}|S|^{4}+\frac{\omega}{2}|\mathcal{H}|^{2}\phi^{2}+\frac{\kappa}{2}\,\phi^{2}|S|^{2}\nonumber \\
 & +\lambda_{3}\,|\mathcal{H}|^{2}|S|^{2}+\lambda_{4}\,|\mathcal{H}^{\dagger}S|^{2}+\{\frac{\lambda_{5}}{2}(\mathcal{H}^{\dagger}S)^{2}+h.c.\}.\label{L}
\end{align}

\begin{figure}[h]
\begin{centering}
\includegraphics[width=0.33\textwidth]{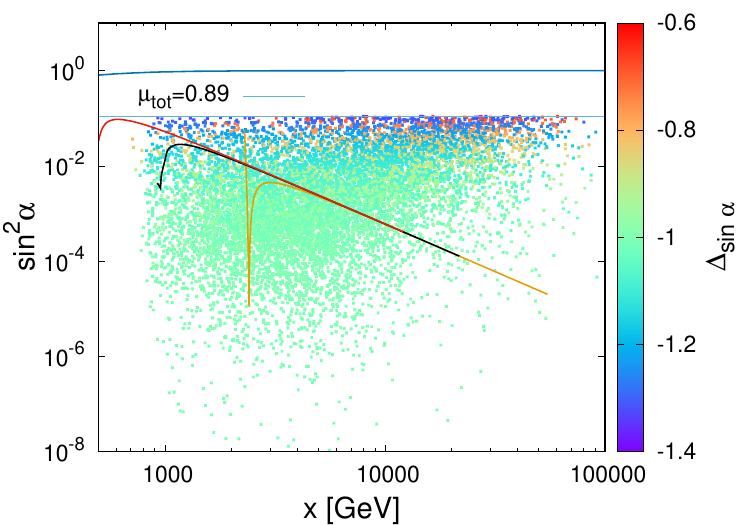}~\includegraphics[width=0.33\textwidth]{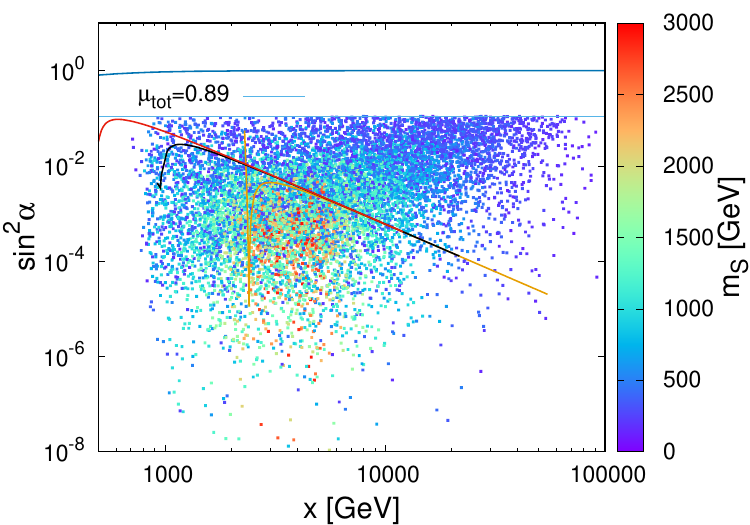}~\includegraphics[width=0.33\textwidth]{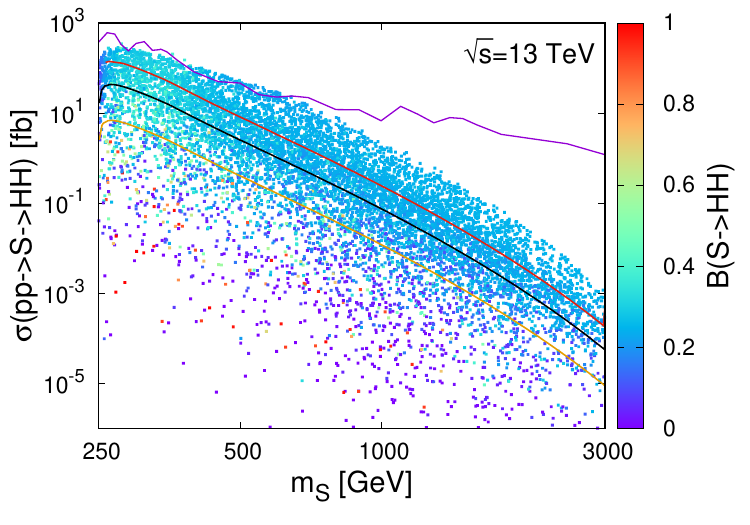}\\
 \includegraphics[width=0.33\textwidth]{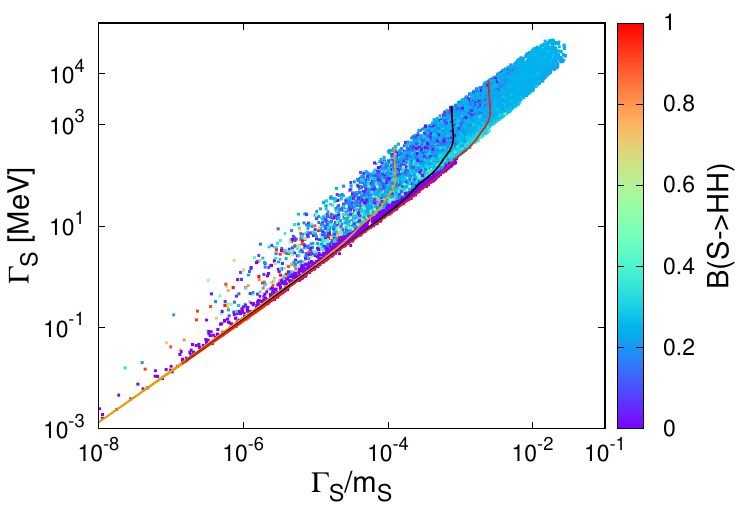}~\includegraphics[width=0.33\textwidth]{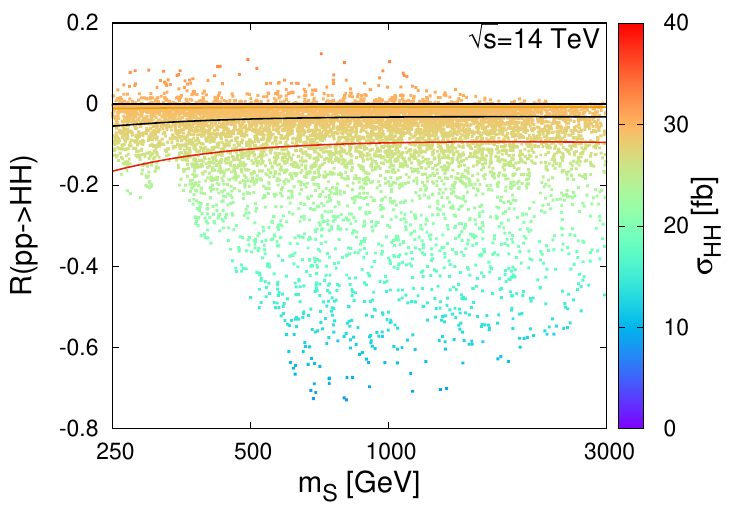}~\includegraphics[width=0.33\textwidth]{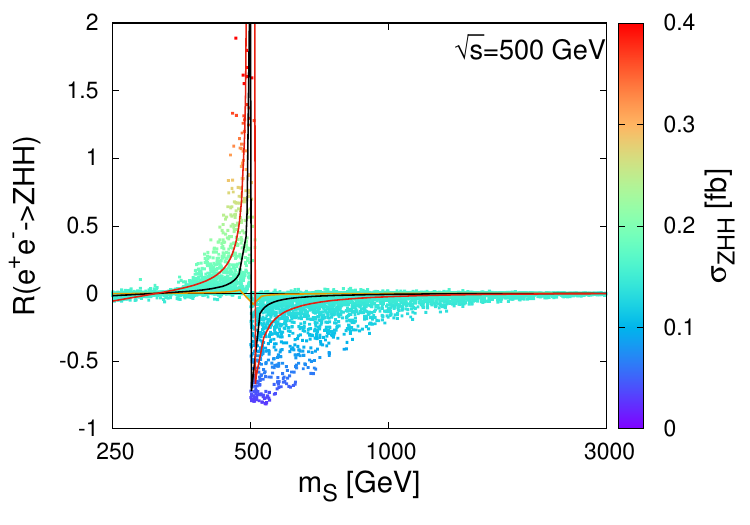} 
\par\end{centering}
\caption{In the upper range, the mixing ($s_{\alpha}^{2}$) versus the singlet
VEV $x$, where the palette shows the relative radiative contribution
to mixing $\Delta_{\sin\alpha}$ (left) and the heavy scalar mass
(middle). The upper blue line corresponds to the tree-level value
of the mixing [$(s_{\alpha}^{(0)})^{2}=x^{2}/(\upsilon^{2}+x^{2})$]
and $\mu_{tot}=0.89$ represents the experimental bound on the Higgs
signal strength in the absence of non-SM Higgs decay modes. The orange,
black and purple lines correspond to the toy model with the parameters
\{$N_{1}=6~,\alpha_{1}=\beta_{1}=0.2$\}, \{$N_{2}=12~,\alpha_{2}=\beta_{2}=0.5$\}
and \{$N_{3}=24~,\alpha_{3}=\beta_{3}=0.9$\}, respectively. In the
top-right panel, we show the resonant di-Higgs production cross section
via the heavy resonance $S$ at $\sqrt{s}=13~\mathrm{TeV}$, compared
to the combination of the recent ATLAS measurements~\cite{ATLAS:2021nps}
(purple curve). Here, the resonant di-Higgs production cross section
estimation was based on the heavy Higgs production cross section values
given in~\cite{HevayHiggs}. In the lower range, we show in the left
panel the total decay width of the heavy resonance versus the ratio
$\Gamma_{S}/m_{S}$, where the palette shows its di-Higgs decay branching
ratio. In the middle and right panels, we show the relative enhancement
(\ref{HH}) for the processes $pp\rightarrow HH@LHC14$ and $e^{-}e^{+}\rightarrow ZHH@ILC500$,
where the palette shows the cross section values in $\mathrm{fb}$.}
\label{fig:1l} 
\end{figure}

In our analysis, we consider the model free parameters to be lying
in the ranges 
\begin{equation}
\begin{array}{c}
x<10^{6}\,\mathrm{GeV},\,y_{i}^{2},\,\left|\lambda_{i}\right|<4\pi,\,M_{DM}<3\,\mathrm{TeV}\end{array}\label{eq:free}
\end{equation}
where $\lambda_{i}$ denotes all the quartic couplings in (\ref{L}).
In Fig.~\ref{fig:1l}, we show many observables that represent either
the relevant constraints on the model or some predictions for current/future
colliders. In order to have an idea about the radiative corrections
effects, we compare our SI-scotogenic results with a toy model, where
the SM is extended by the singlet scalar $\phi$ and a single bosonic
field with with the multiplicity $N_{i}$ and the field dependent
mass $m_{i}^{2}=\frac{1}{2}(\alpha_{i}h^{2}+\beta_{i}\phi^{2})$.
The toy model free parameters \{$N_{i}$,~$\alpha_{i}$ and $\beta_{i}$\}
are constrained by the PRMH requirements [(\ref{dw}), (\ref{cond})
and $\mu_{tot}>0.89$ at 95~\% CL~\cite{ATLAS:2016neq}]; and
the heavy scalar with a mass $m_{H}<m_{S}\leq3\,\mathrm{TeV}$.

One has to mention that for the upper panels range in Fig.~\ref{fig:1l},
we used 10K benchmark points (BPs) and considered many theoretical
and experimental constraints such as the vacuum stability, perturbativity,
perturbative unitarity, electroweak precision tests, the di-photon
Higgs decay, the Higgs invisible decay when applicable, the Higgs
total decay width measurement~\cite{ATLAS:2018jym}, the implications
from negative searches for neutralinos and charginos in supersymmetric
models on the inert masses, the bounds on DM nucleon scattering cross
section from DD experiments (Xenon~1T~\cite{Aprile:2017iyp}); and
the Higgs signal strength at the LHC $\mu_{{\rm tot}}\geq0.89$~\cite{ATLAS:2016neq}.
For the lower panels range, we omitted the BPs that are excluded by
the negative searches for a heavy resonance in the channels $pp\rightarrow S\rightarrow\tau\tau,\,ZZ,\,WW$
and by the negative searches on the resonant di-Higgs production $pp\rightarrow S\rightarrow HH$
via the different channel $HH\rightarrow b\bar{b}\tau\tau,\,b\bar{b}b\bar{b},\,b\bar{b}\gamma\gamma\,$
as mentioned previously. These constraints exclude only 5.35\% of
the BPs used in the upper panels in Fig.~\ref{fig:1l}. Indeed, there
are other relevant constraints to this model such as neutrinos oscillation
data, DM relic density and and the lepton flavor violating processes.
These constraints are not considered here since we interested on the
parameters and constraints that are relevant to the radiative effects
on the Higgs sector.

From Fig.~\ref{fig:1l}, one can learn many conclusions. A PRMH scenario
is viable for a large parameters space, where the radiative corrections
can give rise to the Higgs mass and simultaneously push the scalar
mixing to be in agreement with the total Higgs strength bound~\cite{ATLAS:2016neq}.
For instance, for heavy scalar masses below $3~\mathrm{TeV}$, the
one-loop quartic couplings $\lambda_{h}^{1-\ell}$ and $\omega^{1-\ell}$
are not practically constrained by the perturbativity since they are
lying in the ranges $[0.07,2.5]$ and $[-1,2]$, respectively. However,
the singlet one-loop quartic coupling $10^{-5}<\lambda_{\phi}^{1-\ell}\leq4\pi$,
together with the previous requirements make the singlet scalar VEV
lies in the range $600~\mathrm{GeV}<x<100~\mathrm{TeV}$. Here, the
fact that the heavy scalar is barely constrained by the recent RUN-II
measurements of ATLAS with 139 $\mathrm{fb}^{-1}$~\cite{ATLAS:2020zms,ATLAS:2020tlo,ATLAS:2021nps},
and CMS with 137 $\mathrm{fb}^{-1}$~\cite{CMS:2021klu}, this scenario
would be within the reach of the coming analysis.

One has to notice that the total decay width of the heavy scalar is
much smaller than its mass for most of the viable parameters space,
and therefore the narrow width approximation used to estimate the
resonant di-Higgs production cross section is justified. In addition,
the cross section of the non-resonant di-Higgs production at the LHC
$pp\rightarrow HH@LHC14$ is reduced (by up to 75\%) for the majority
of the parameters space, while it is enhanced for few BPs by less
than 10\%. For the Z-associated di-Higgs production at the ILC $e^{-}e^{+}\rightarrow ZHH@ILC500$,
the cross section is mainly enhanced for $m_{S}<500~\mathrm{GeV}$,
and reduced for larger values $m_{S}>0$. In this setup, the enhancement/suppression
is maximal around $m_{S}\sim\sqrt{s}=500\,\mathrm{GeV}$, since it
is not a numerical mis-estimation of the cross section due to the
Breit-Wigner corrected propagators used in (\ref{eq:coef}). In case
where the measured Z-associated di-Higgs production is reduced (increased)
with respect to the SM by less than 30\% (more than 100\%), the heavy
scalar mass is $m_{S}\gtrsim500~\mathrm{GeV}$ ($m_{S}\lesssim500~\mathrm{GeV}$).
For completeness, one has to mention that the BPs in Fig.~\ref{fig:1l}
are in agreement with DM constraints such as the DD bounds and the
relic density. Here, we enforced the relic density to be $\Omega_{N_{1}}h^{2}>0.12$
due to the annihilation channels $N_{1}N_{1}\rightarrow VV,HH,HS,SS,f\bar{f}$,
where the contribution of the channel $N_{1}N_{1}\rightarrow\ell_{\alpha}\ell_{\beta},\nu_{\alpha}\bar{\nu}_{\beta}$
to the annihilation cross section would relax the relic density to
match the measured value~\cite{Soualah:2021xbn}, $\Omega_{DM}h^{2}=0.120\pm0.001$~\cite{Aghanim:2018eyx}.

The idea of the Higgs as a PGB in a SI framework has been discussed
in~\cite{Foot:2007as}. In addition to the EWSB details discussion,
the authors had shown that the light CP-even mass could exceed the
Higgs mass bound (then, $m_{H}>114~\mathrm{GeV}$). They considered
two phenomenologically consistent models to validate this possibility.
Although in SI models, it has been shown that the slow-roll inflation
can be achieved by adding a extra VEVless singlet real scalar that
is coupled non-minimally to the gravity. This real field singlet inflationary
model does not suffer from a unitarity breakdown at a scale below
or comparable to the inflation scale~\cite{Khoze:2013uia}. Here,
the singlet field that is responsible for inflation can be also a
viable DM candidate. In a non SI model~\cite{Aravind:2015xst} that
is similar to our illustrative example, where fermionic DM has been
addressed and the EWSB is assisted by a real scalar singlet, it has
been shown that the inflaton could either be the Higgs boson or the
singlet scalar, and slow-roll inflation can be realized via a non-minimal
coupling to gravity. This tells us that achieving a successful slow-roll
inflation within the PRMH scenario deserves an extensive investigation
to define the viable parameters space region(s).

\section{Conclusion\label{sec:conc}}

In this work, we have shown that the PRMH scenario within the scale
invariance approach is possible; where we have derived the condition
the required conditions to be fulfilled by the masses, couplings and
multiplicities of the new fields added to the SM. We have described
also the experimental constraints that are coming for the recent ATLAS
and CMS measurements on the Higgs properties and the negative searches
of heavy resonances. Significant part of the parameters space makes
this scenario in a good agreement with the data. We have proven that
to avoid the constraints from the total Higgs signal strength modifier~\cite{ATLAS:2016neq},
the radiative corrections that give rise to the Higgs mass must be
considered in order to push the singlet-doublet scalar mixing to lie
in the experimentally allowed region. This leads to non-negligible
values for the triple scalar couplings $\lambda_{HHH}$ and $\lambda_{HHS}$,
that are strictly vanishing at tree-level. Thus, the PRMH scenario
is very sensible to the resonant di-Higgs production $pp\rightarrow S\rightarrow HH$~\cite{ATLAS:2021nps},
as well the non-resonant ones $pp\rightarrow HH$ and $e^{-}e^{+}\rightarrow ZHH$.
We have considered the SI-scotogenic model~\cite{Ahriche:2016cio}
as an illustrative example, where we have checked different experimental
constraints and given some predictions about (Z-associated) di-Higgs
production at ($ILC500$) $LHC14$. The PRMH scenario looks interesting
since many physical observables are all triggered together by the
radiative effects, and therefore, other aspects should be investigated
within this approach, such as the electroweak phase transition (EWPT)
strength, gravitational waves produced during the EWPT in addition
to the different collider signatures that are relevant to the triple
scalar couplings.

\appendix

\section{Scalar Triple Couplings}

\label{app}

The triple scalar couplings $\lambda_{HHH}$ and $\lambda_{HHS}$
can be defined as combination of the third derivative of the scalar
potential after the EWSB. For example, the coupling $\lambda_{HDD}$
can be estimated as~\cite{Ahriche:2013vqa} 
\begin{equation}
\begin{array}{c}
\lambda_{HHH}=\left.\left\{ c_{\alpha}^{3}V_{h,h,h}^{1-\ell}-3s_{\alpha}c_{\alpha}^{2}V_{h,h,\phi}^{1-\ell}+3s_{\alpha}^{2}c_{\alpha}V_{h,\phi,\phi}^{1-\ell}-s_{\alpha}^{3}V_{\phi,\phi,\phi}^{1-\ell}\right\} \right|_{h=\upsilon,\,\phi=x},\\
\lambda_{HHS}=\left.\left\{ c_{\alpha}^{2}s_{\alpha}V_{h,h,h}^{1-\ell}+(c_{\alpha}^{3}-2c_{\alpha}s_{\alpha}^{2})V_{h,h,\phi}^{1-\ell}+(s_{\alpha}^{3}-2c_{\alpha}^{2}s_{\alpha})V_{h,\phi,\phi}^{1-\ell}+c_{\alpha}s_{\alpha}^{2}V_{\phi,\phi,\phi}^{1-\ell}\right\} \right|_{h=\upsilon,\,\phi=x},
\end{array}\label{L3}
\end{equation}
with $V_{x,y,z}^{1-\ell}\equiv\partial^{3}V^{1-\ell}/\partial x\partial y\partial z$.
The reason that these couplings vanish at tree-level (i.e., by considering
the tree-level potential and the mixing $s_{\alpha}^{(0)}$); is due
to the tree-level vacuum structure of all SI SM extensions, where
the EWSB is assisted by the real singlet scalar $\phi$. The one-loop
couplings in (\ref{L3}) can be estimated by considering the one-loop
effective potential (\ref{eq:V1l}) and the tree-level mixing $s_{\alpha}^{(0)}$.
However, one can obtain more precise values by doing some re-summation.
Here, we will use a resummed estimation of the couplings (\ref{L3})
by taking into account the one-loop effective potential (\ref{eq:V1l})
and the one-loop mixing $s_{\alpha}^{(1)}$ instead of the tree-level
mixing $s_{\alpha}^{(0)}$. We found that the resummed one-loop values
for (\ref{L3}) are significantly different from zero; and they are
fully triggered by quantum corrections.

\subsection*{Acknowledgements}

This work is funded by the University of Sharjah under the research
projects No 21021430100 ``\textit{Extended Higgs Sectors at Colliders:
Constraints \& Predictions}'' and No 21021430107 ``\textit{Hunting
for New Physics at Colliders}''.


\end{document}